\newcommand{\dndt}{\mbox{${\rm d}N/{\rm d}t$}}

\newcommand{\tdis}{\mbox{$t_{\rm dis}$}}

\newcommand{\tdistid}{\mbox{$t^{\rm tid}_{\rm dis}$}}
\newcommand{\tdisgmc}{\mbox{$t^{\rm GMC}_{\rm dis}$}}
\newcommand{\trel}{\mbox{$t_{\rm rel}$}}
\newcommand{\kms}{\mbox{${\rm km}\,{\rm s}^{-1}$}}
\newcommand{\msun}{\mbox{$M_\odot$}}
\newcommand{\mlim}{\mbox{$M_{\rm lim}$}}
\newcommand{\mmin}{\mbox{$M_{\rm min}$}}

\newcommand{\mvlim}{\mbox{$M_{V,{\rm lim}}$}}
\newcommand{\mvssp}{\mbox{$M_{V}^{\rm SSP}$}}

\newcommand{\rhon}{\mbox{$\rho_{\rm n}$}}

\newcommand{\rg}{\mbox{$R_{\rm G}$}}
\newcommand{\rh}{\mbox{$r_{\rm h}$}}

\newcommand{\mc}{\mbox{$M_{\rm c}$}}
\newcommand{\vg}{\mbox{$V_{\rm G}$}}

\newcommand{\msunpc}{\mbox{$M_\odot\,{\rm pc}^{-3}$}}

\documentclass{iaus}
\usepackage{graphicx,natbib}

\title[Star cluster life-times] 
{Star cluster life-times: dependence on mass, radius and environment}

\author[Gieles, Lamers \& Baumgardt]   
{Mark Gieles$^1$,
 Henny~J.~G.~L.~M. Lamers$^2$ \and
Holger Baumgardt$^3$}
 
\affiliation{$^1$European Southern Observatory, Casilla 19001, Santiago 19, Chile\break
email: mgieles@eso.org\\[\affilskip]
$^2$Astronomical Institute, Utrecht University, 
  Princetonplein 5, \break 3584 CC Utrecht, The Netherlands  \break email: lamers@astro.uu.nl \\[\affilskip]
$^3$Argelander Institut f\"ur Astronomie, Universit\"at Bonn, Auf dem H\"ugel 71, Bonn, Germany 
 \break email: holger@astro.uni-bonn.de
  }

\pubyear{2007}
\volume{246}  
\pagerange{001--005}
\date{?? and in revised form ??}
\jname{Dynamical Evolution of Dense Stellar Systems}
\editors{E. Vesperini, M. Giersz \& A. Sills, eds.}
\begin{document}

\maketitle

\begin{abstract}
The dissolution time (\tdis) of clusters in a tidal field does not
scale with the ``classical'' expression for the relaxation
time. First, the scaling with $N$, and hence cluster mass, is
shallower due to the finite escape time of stars. Secondly, the cluster
half-mass radius is of little importance. This is due to a balance
between the relative tidal field strength and internal relaxation,
which have an opposite effect on \tdis, but of similar magnitude.
When external perturbations, such as encounters with giant molecular
clouds (GMC) are important, \tdis\ for an individual cluster depends
strongly on radius. The mean dissolution time for a population of
clusters, however, scales in the same way with mass as for the tidal
field, due to the weak dependence of radius on mass. The environmental
parameters that determine \tdis\ are the tidal field strength and the
density of molecular gas. We compare the empirically derived \tdis\ of
clusters in six galaxies to theoretical predictions and argue that
encounters with GMCs are the dominant destruction mechanism.  Finally,
we discuss a number of pitfalls in the derivations of \tdis\ from
observations, such as incompleteness, with the cluster system of the
SMC as particular example.
\keywords{globular clusters: general, open clusters and associations: general, stellar dynamics, methods: n-body simulations}
\end{abstract}

\section{Theoretical predictions of cluster dissolution}
\label{sec:}

\subsection{Dynamical evolution in a tidal field}
\label{ssec:tidal}
Simulations of star clusters dissolving in a tidal field have shown that the dissolution time (\tdistid) scales with the relaxation time (\trel) as $\tdistid\propto t^{0.75}_{\rm rel}$ \citep{2001MNRAS.325.1323B, 2003MNRAS.340..227B}. This non-linear dependence on \trel\ is due to the finite escape time through one of the Lagrange points \citep{2000MNRAS.318..753F}. The dependence on $N$, or cluster mass (\mc), can be approximated as $\tdistid\propto\mc^{0.62}$, which is accurate for $10^2\lesssim N\lesssim10^7$ \citep{2005A&A...429..173L}.
  The half-mass radius (\rh) of the cluster does not enter in the results, since it is assumed that clusters are initially ``Roche lobe" filling, which implies $\mc\propto r^3_{\rm h}$, i.e. a constant crossing time.

The assumption of Roche lobe filling clusters is computationally attractive since it avoids having \rh\ as an extra parameter. However,  observations of (young) extra-galactic star clusters show that the dependence of \rh\ on \mc\ and galactocentric distance (\rg) is considerably weaker ($\rh\propto M^{0.1}\,R^{\,0.1}_{\rm G}$) than the Roche lobe filling relation ($\rh\propto M^{1/3}\,R_{\rm G}^{\,2/3}$) \citep{2004A&A...416..537L, 2007A&A...469..925S}, implying that massive clusters at large \rg\ are initially underfilling their Roche lobe.

\begin{figure}
\center
 \includegraphics[width=7cm]{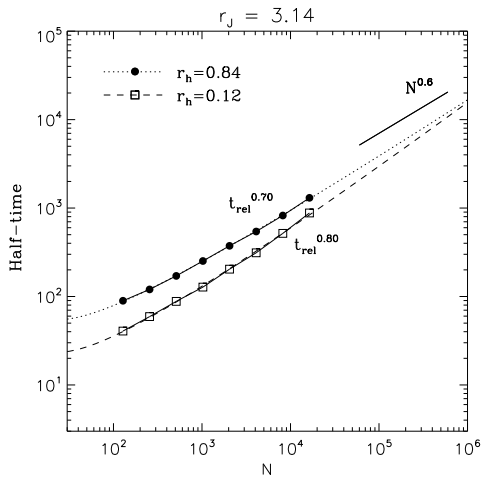}
  \caption{Half-mass time as found from $N$-body simulations of clusters dissolving in a tidal field. The filled circles represent clusters that initially fill their Roche lobe. The open squares are the results of runs where \rh\ was seven times smaller.}\label{fig:01}
\end{figure}

\cite{gieles07} simulated clusters with varying initial \rh\ in a tidal field to quantify the importance of \rh. Figure~\ref{fig:01} shows the results of \tdistid\ for two sets of clusters with different initial \rh. The filled circles are for clusters that started tidally limited and the open squares are for runs where the initial \rh\ was a factor seven smaller. 
The difference in \tdis\ are within a factor two, while the
``classical" expression of \trel\ predicts a factor
$7^{3/2}\simeq20$. The reason that \tdis\ depends so little on \rh\
can be understood intuitively: for smaller clusters the tidal field is
less important, but the dynamical evolution is faster. These effects
happen to balance and result in almost no dependence on \rh. {\it The
crossing of the lines around $N\simeq10^6$ implies that for globular
clusters \tdistid\ is completely independent of \rh.}
 
This somewhat surprising result means that we can use the \rh\ independent results for \tdis\ of tidally limited clusters  \citep{2003MNRAS.340..227B} as a general result for \tdistid\ for clusters of different \rh:

\begin{equation}
\frac{\tdistid}{\rm Gyr}=1.0\,\left(\frac{\mc}{10^4\,\msun}\right)^{0.62}\,\frac{\rg}{\vg}\frac{220\,\kms}{\rm kpc}.
\label{eq:1}
\end{equation}
From this it follows that a cluster with $\mc=10^4\,\msun$ in the solar neighbourhood would dissolve in approximately 8 Gyr due to tidal field. This is much longer than the empirically derived value of $1.3\,$Gyr \citep{2005A&A...441..117L}, implying that there are additional disruptive effects that shorten the life-time of clusters.

\subsection{External perturbations: disruption by giant molecular clouds}
\label{ssec:ext}

It has long been suspected that encounters with giant molecular clouds (GMCs) shorten the life-times of clusters (e.g. \citealt{1980A&A....88..360V}). \cite{2006MNRAS.371..793G} studied this effect using $N$-body simulations and found that \tdis\ due to GMC encounters (\tdisgmc) can be expressed in cluster properties and average molecular gas density (\rhon) as

\begin{equation}
\frac{\tdisgmc}{\rm Gyr}=2.0\,\left(\frac{0.03\,\msunpc}{\rhon}\right)\left(\frac{\mc}{10^4\,\msun}\right)\,\left(\frac{3.75\,{\rm pc}}{\rh}\,\right)^3. 
\label{eq:2}
\end{equation}
The scaling of \tdisgmc\ with cluster density ($\mc/r^3_{\rm h}$) combined with the observed weak dependence of \rh\ on \mc,  $\rh \propto\mc^{0.13}$, results in a similar scaling of the mean \tdisgmc\ with \mc\  as found for \tdistid\ , i.e. $\propto\mc^{0.6}$ (\ref{eq:1}). 

For the solar neighbourhood ($\rhon\simeq0.03\,\msunpc$) $\tdisgmc=2\,$Gyr, which combined with the tidal field (\ref{eq:1}) nicely explains the emperically derived \tdis\ of 1.3\,Gyr and the observed age distribution of clusters in the solar neighbourhood \citep{2006A&A...455L..17L}.

From (\ref{eq:1}) and (\ref{eq:2}) we see that the predicted \tdis\ scales with the tidal field strength (\rg/\vg) and the inverse of the molecular gas density (1/\rhon). In table~\ref{tab:1} we give values for these parameters for six galaxies, combined with predictions for $t_4$. The values for \rhon\ are taken from \citet{2006MNRAS.371..793G} (and references therein), \citet{2004ApJ...602..723H, 2007ApJ...658.1027L} for the solar neighbourhood, M51, M33 and the SMC, respectively. In the next section we compare this to empirically derived values of \tdis.

\section{Comparison to observations}

\subsection{Empirically derived \tdis\ values in different galaxies}
Under the assumption that \tdis\ scales with \mc, \citet{2003MNRAS.338..717B} (BL03) introduced an empirical disruption law: $\tdis=t_4\,(\mc/10^4\,\msun)^\gamma$. The value of $t_4$ and $\gamma$ can be derived from the age and mass distributions (see BL03 for details). BL03 found a mean $\gamma$ of  $\bar\gamma=0.62$, agreeing nicely with (\ref{eq:1}) and (\ref{eq:2}), and values for $t_4$ ranging from $\sim100\,$Myr to $\sim8\,$Gyr. We summarise values of $t_4$ of clusters in six different galaxies taken from more recent  literature in table~\ref{tab:1}.

Note that   $\rg/\vg$ and $1/\rhon$ roughly increase with increasing $t_4$. The variation in $\rg/\vg$ is too small to explain the variation in $t_4$, which implies that in the galaxies with short $t_4$ the disruption is dominated by GMC encounters. {\it From Table~\ref{tab:1} we see that the decreasing trend in the emperical $t_4$ can be explained by increasing gas density and increasing tidal field strength.}

\begin{table}\def~{\hphantom{0}}
  \begin{center}
  \caption{Columns 1-3: Estimates of tidal field strength, molecular gas densities and resulting predictions for $t_4$, the \tdis\ of a cluster with an initial $\mc=10^4\,\msun$. Column 4: emperically derived values of $t_4$ are given, taken from: $^1$\cite{2005A&A...441..949G}; $^2$\cite{2005A&A...429..173L}; $^3$\cite{2005A&A...441..117L}; $^4$Parmentier \& de Grijs (2007); $^5$\cite{2004PASP..116..497K}; $^6$\cite{2003MNRAS.338..717B}.}
  \label{tab:1}
  \begin{tabular}{lcccc}\hline
      Galaxy  			&   Tidal field  	& Molecular gas density       & Predicted $t_4$& Observed $t_4$	\\
                     		&  $\rg/\vg$  [Myr] 	& $\rhon$ [$10^{-3}\,\msunpc$] & [Gyr]         &   [Gyr] 		 \\\hline
       M51$^1$   		&  ~~10		        & 450		               & ~0.13          &  ~Ê0.1		\\
       M33$^2$   		&  ~~15			&  ~25			       & ~~1.4           &  ~Ê0.6		\\
       Solar neighbourhood$^3$	&  ~~35			&  ~30			       & ~~1.6	       &  ~1.3			\\
       LMC$^4$   		&  ~~30			&  ~~-			       & $<6.6$	       &  ~$>$1			\\
       NGC6822$^5$  		&  $\sim35$		&  ~~-			       & $<7.7$	       &  $\sim4$		   \\
       SMC$^6$  		&  ~~40			& 0.5			       & ~~8.2	       &  ~Ê~8		        \\\hline
    \end{tabular}
 \end{center}
\vspace{-0.2cm}
\end{table}

\subsection{The clusters of the SMC}

\begin{figure}
\center
 \includegraphics[width=8cm]{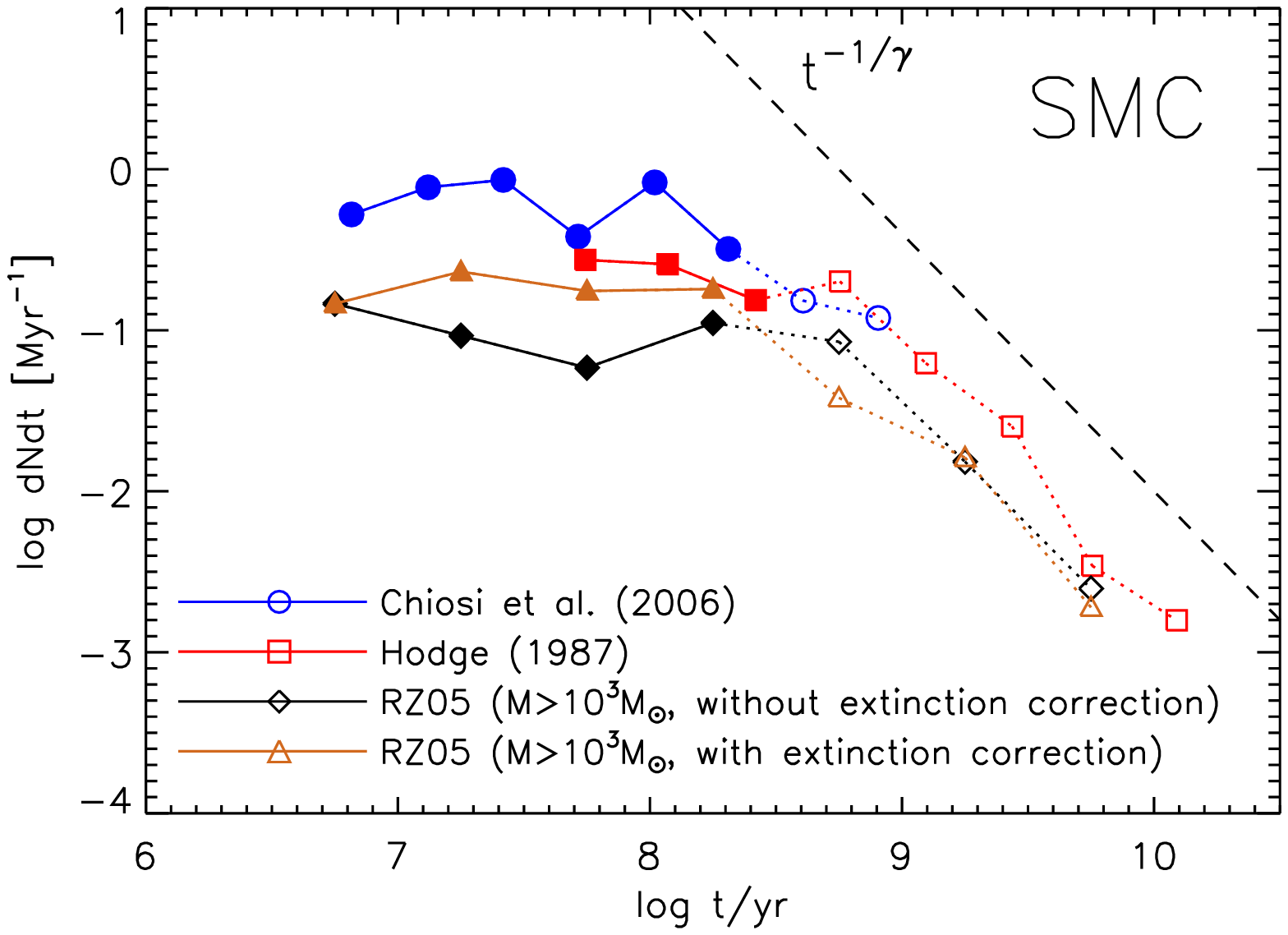}
  \caption{The age distribution (\dndt) of star clusters in the SMC as found in different  studies in literature. The data set of \cite{2005AJ....129.2701R} (RZ05) is very incomplete for low mass clusters at old ages \citep{2007arXiv0706.1202G}, so a mass cut at $10^3\,\msun$ was applied. The general trend found in these studies is that \dndt\ is flat up to an age of $0.3-1\times10^{9}\,$yr and then it declines as $t^{-1.7}$. The dashed line is the predicted slope for \dndt\ at old ages when $\tdis\propto\mc^\gamma$, with $\gamma=0.62$. }
  \label{fig:2}
\end{figure}

A lot of attention has gone recently to the age distribution (\dndt) of clusters in the SMC. \cite{2005AJ....129.2701R} (RZ05) found that \dndt\ is roughly declining as $t^{-1}$, which  \cite{2006ApJ...650L.111C} explain by mass independent cluster disruption\footnote{In fact the authors call their disruption model ``infant mortality", but we prefer to reserve this term for the dissolution of clusters due to gas expulsion. In addition, 3 Gyr old clusters have survived 25\% of a Hubble time, so they are not really infant anymore.} removing 90\% of the clusters each age dex. \cite{2007arXiv0706.1202G} showed that the decline is caused by incompleteness and that the \dndt\ is flat in the first $\sim1\,$Gyr when using a mass limited sample. The \dndt\ based on ages which are derived from extinction corrected colours starts declining  a bit earlier than the one  based on uncorrected colours (figure~\ref{fig:2}). However, the general shape is similar to that found by other authors: a flat part in the first $0.3-1.0\,$Gyr (recently reconfirmed by \citealt{2007arXiv0709.3781D}) and then a steep decline ($\propto t^{-1.7}$).  When $\tdis\propto\mc^{\gamma}$, then the \dndt\ at old ages declines as $t^{-1/\gamma}$ for both mass and magnitude limited samples (BL03). The decline of $t^{-1.7}$ implies $\gamma\simeq0.6$, in agreement with the theoretical predictions (\ref{eq:1} and \ref{eq:2}).

\subsection{Selection effects and biases: a cautionary note}
\begin{figure*}
\centering
 \includegraphics[width=10.5cm]{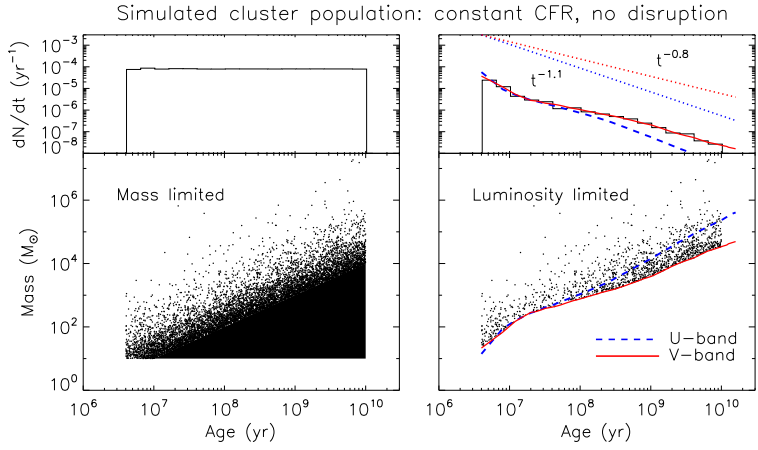}
  \caption{Simulated ages and masses of a cluster population that has formed with a constant cluster formation rate (CFR) and with a power-law CIMF ($N\propto M^{-2}$). In the left panels we show the result of mass limited sample, with $\mlim=10\,\msun$. In the right panels we assume that the sample is magnitude limited, with $\mvlim=-4.5$. The limiting mass due to a magnitude limit and the resulting prediction for \dndt\ of a magnitude limited sample are shown as full lines (red). The prediction for a $U$-band limit ($M_U=-5$) is shown as dashed lines (blue). The dotted lines show power-law approximations for the predicted shapes of \dndt.}
  
  \label{fig:3}
\end{figure*}

Observed cluster samples are always heavily affected by the detection limit, causing the minimum observable cluster mass (\mmin) to increase with age, due to the fading of clusters. To illustrate this effect we create an artificial cluster population with a constant cluster formation rate (CFR) and with a power-law CIMF with index $-2$. In the left panels of figure~\ref{fig:3} we show the ages and masses (bottom) and the corresponding \dndt\  (top) when the sample is mass limited. The \dndt\ is flat which is the result of the constant CFR we put it. In the right panel we remove the clusters which are fainter than $M_V=-4.5$. The mass of a cluster at the detection limit, $\mmin(t)$, increases with age as $\mmin(t)\propto0.4\,\mvssp(t)$, where $\mvssp(t)$ is the evolution of $M_V$ with age from an SSP model. For a power-law CIMF with index $-2$, the resulting \dndt\ scales with \mmin\ as $\dndt\propto 1/\mmin(t)$  (BL03), which is shown in the top right panel of figure~\ref{fig:3}. 

The detection limit is usually expressed in $M_V$. However, deriving cluster ages from broad band photometry requires the presence of blue filters such as $U$ and $B$. We show the
$\mmin(t)$ for  a $U$-band detection limit of $M_U=-5$ as a dashed line in the age {\it vs.} mass diagram. The resulting \dndt\ (shown as a dashed line in the top right panel) declines approximately as $\dndt\propto t^{-1.1}$, i.e. steeper than the $V$-band prediction. {\it It is of vital importance to understand the effect of incompleteness in different filters before a disruption analyses can be done based on the slope of the \dndt\ distribution.}

%




\end{document}